\documentclass[5p,times,preprint,twocolumn]{elsarticle}

\usepackage{amssymb}
\usepackage{mathrsfs}

\usepackage[figuresright]{rotating}
\usepackage{booktabs}
\usepackage{xcolor}
\usepackage{multirow}
\usepackage{graphicx}
\usepackage{amssymb}
\usepackage{amsmath}
\usepackage{tabularx}
\usepackage{hyperref}

\begin{document}

\begin{frontmatter}




\title{Federative ischemic stroke segmentation as alternative to overcome domain-shift multi-institution challenges}

\journal{Radiology: Artificial Intelligence}

\author[bivl2ab]{Edgar Rangel}
\ead{edgar.rangel@correo.uis.edu.co}
\author[bivl2ab]{Fabio Mart\'inez\corref{cor}}
\ead{famarcar@saber.uis.edu.co}

\cortext[cor]{Corresponding author}

\affiliation[bivl2ab]{organization={Biomedical Imaging, Vision and Learning Laboratory (BIVL$^2$ab)},
    addressline={Universidad Industrial de Santander},
    country={Colombia}}

\begin{abstract}
Stroke is the second leading cause of death and the third leading cause of disability worldwide. Clinical guidelines establish diffusion resonance imaging (DWI, ADC) as the standard for localizing, characterizing, and measuring infarct volume, enabling treatment support and prognosis. Nonetheless, such lesion analysis is highly variable due to different patient demographics, scanner vendors, and expert annotations. Computational support approaches have been key to helping with the localization and segmentation of lesions. However, these strategies are dedicated solutions that learn patterns from only one institution, lacking the variability to generalize geometrical lesions shape models. Even worse, many clinical centers lack sufficient labeled samples to adjust these dedicated solutions. This work developed a collaborative framework for segmenting ischemic stroke lesions in DWI sequences by sharing knowledge from deep center-independent representations. From 14 emulated healthcare centers with 2031 studies, the FedAvg model achieved a general DSC of $0.71 \pm 0.24$, AVD of $5.29 \pm 22.74$, ALD of $2.16 \pm 3.60$ and LF1 of $0.70 \pm 0.26$ over all centers, outperforming both the centralized and other federated rules. Interestingly, the model demonstrated strong generalization properties, showing uniform performance across different lesion categories and reliable performance in out-of-distribution centers (with DSC of $0.64 \pm 0.29$ and AVD of $4.44 \pm 8.74$ without any additional training).
\end{abstract}


\end{frontmatter}


\begin{figure*}
    \centering
    \includegraphics[width=0.95\textwidth]{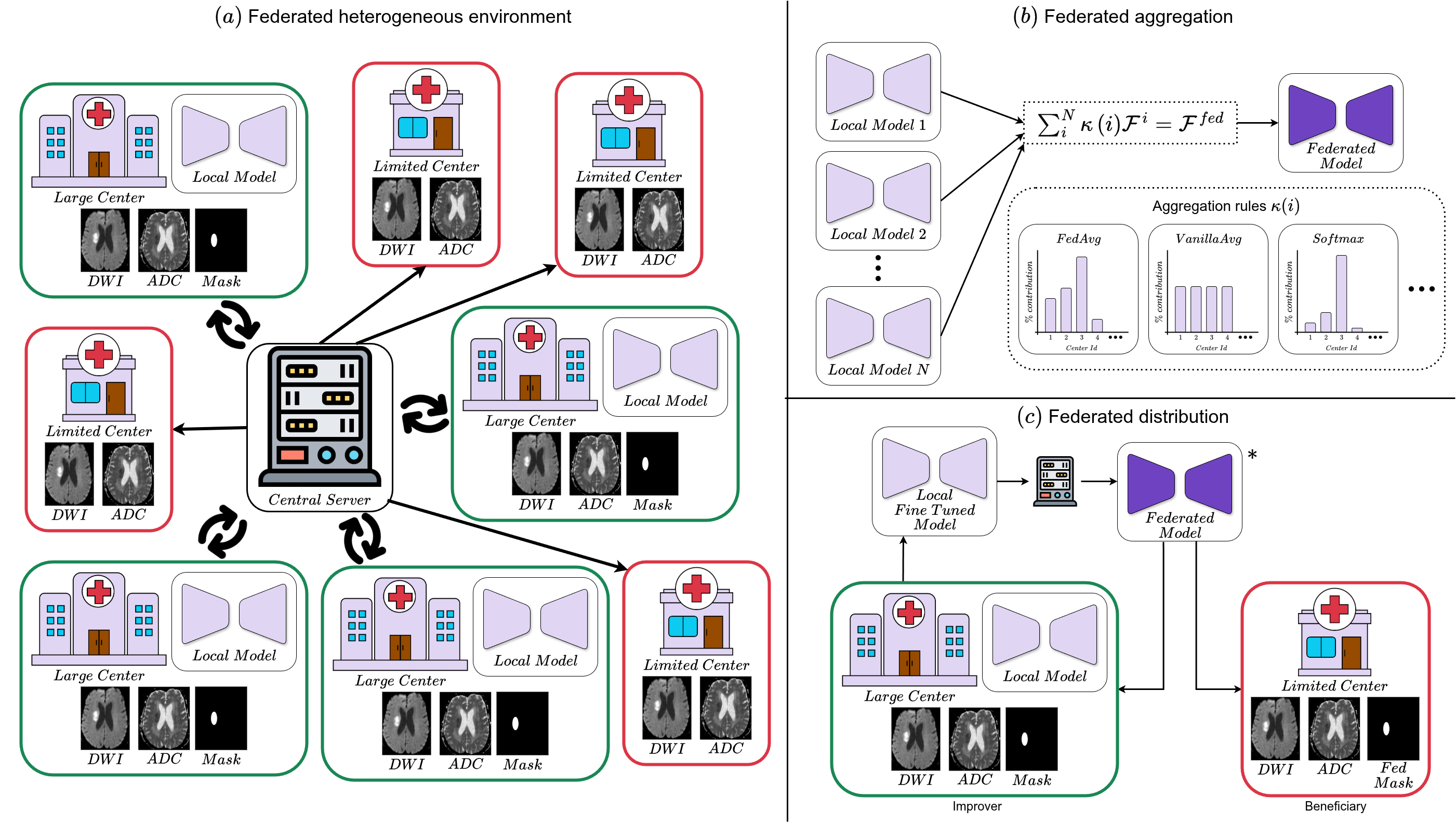}
    \caption{The proposed approach pipeline. In (a) there are multiple heterogeneous large and limited centers with and without training capabilities, respectively. Then in (b), the large centers deliver their models to a central server which carries out a federated aggregation rule to generate a single federated model. Lastly, in (c) the federated model is delivered to large centers to fine-tune their models and repeat the process to improve the federated model, while limited centers benefit from the federated representations to generate segmentations and support experts inside the healthcare center.}
    \label{fig:pipeline}
\end{figure*}

\section{Introduction}
Stroke, a disruption in the brain blood flow, is the second leading cause of death and the third leading cause of disability worldwide \cite{feigin2022world}. The ischemic stroke, caused by blood vessel occlusions, represents more than 80\% of all reported stroke cases \cite{tsao2023heart}. According to clinical guidelines, radiologists can discriminate between affected and healthy brain tissue from both Diffusion-Weighted Imaging sequences (DWI) and Apparent Diffusion Coefficient (ADC) parametric maps \cite{powers2019guidelines,goyal2020challenging}. Particularly, ischemic stroke lesions are principally perceived in DWI as hyperintense signals, which are the response of intracellular cytotoxic and water accumulation. Complementary, the lesion shape and localization are confirmed by observation on complementary ADC maps \cite{sperber2023stroke}. In these maps, lesions appear as hypointense signals, calculated by varying the amplitude and duration of bipolar gradients of molecules (\textit{a.k.a} b-values) in DWI between 0 $s/mm^2$ (DWI-B0) and 1000 $s/mm^2$ (DWI-B1000) \cite{le2001diffusion}. Both modalities are key for identifying ischemic infarcts within the critical thrombolytic window, especially when there exists uncertainty of onset symptoms \cite{juan2022improving,nukovic2023neuroimaging}. Nonetheless, such lesion characterization is expert-dependent, which in turn is challenging due to the high variability in lesion shape, size, and location \cite{sperber2023stroke}. Particularly, while neuroradiologists often agree on the lesion's core, there can be significant disagreement regarding lesions boundaries, hindering accurate volume estimation, a critical factor in treatment decisions and outcome prognosis \cite{neumann2009interrater}.

Computational approaches have emerged as a potential alternative to mitigate lesion characterization challenges and support lesion detection over MRI studies \cite{abbasi2023automatic, luo2024deep}. However, these solutions learn from a dedicated set of data, typically from a unique healthcare center, with biased expert annotations \cite{abbasi2023automatic}. Hence, a significant and variable amount of data is required in terms of patient demography, symptoms onset diversity, and even integrating annotations from several radiologists to deal with observation subjectivity \cite{luo2024deep}. A main challenge to achieve such data requirements is an effective collaboration among centers for knowledge sharing, because of limitations related to data privacy and security \cite{yoo2022open}. In fact, there exist multiple regulations (for instance, HIPAA or GDPR legislations) and standards that are constantly updated to guarantee patient privacy, but resulting each time more restrictive for center collaboration \cite{rieke2020future,dhade2024federated}. Even, each institution has proper characteristics regarding data, experts, and computational capabilities, evidencing the necessity to construct personalized or collaborative solutions that can be mapped or adapted, according to the requirements of each center \cite{yoo2022open}. Federated Learning (FL) has emerged as a promising alternative to share computational representations protecting patient privacy and enabling collaborative research across institutions by considering their own characteristics. Particularly, for stroke segmentation some seminal works have evidenced opportunities in this collaborative learning \cite{otalora2022weighting,madrona2023federated,rangel2024federated}. However, such approaches are limited in variability regarding expert annotations and lesions' properties, along with the emulation of heterogeneous independent institutions given the unique source of data.

In this work a federated learning framework was developed to enable collaborative knowledge from ischemic stroke segmentation models. The model showed strong generalization properties across different institutions and supported segmentation in heterogeneous lesions. The key contributions of this works are: 

\begin{itemize}
    \item The study employed 2031 DWI and ADC neuroimages, from 14 emulated healthcare centers, based on three public and two private datasets. The institutions were divided into large and limited centers (with AI training capabilities and out-of-distribution institutions, respectively) to validate the models.
    \item Federate Learning enables a collaborative platform for sharing knowledge to tackle ischemic stroke variability, leveraging heterogeneous scenarios such as patients demography, expert annotations, scanner vendors and lesions morphology. 
    \item The FedAvg model achieved the best performance against the centralized and other federated rules, evidencing a reliable and uniform performance in seen and out-of-distribution institutions, along with several lesion morphologies.
\end{itemize}

\begin{figure*}
    \centering
    \includegraphics[width=0.95\textwidth]{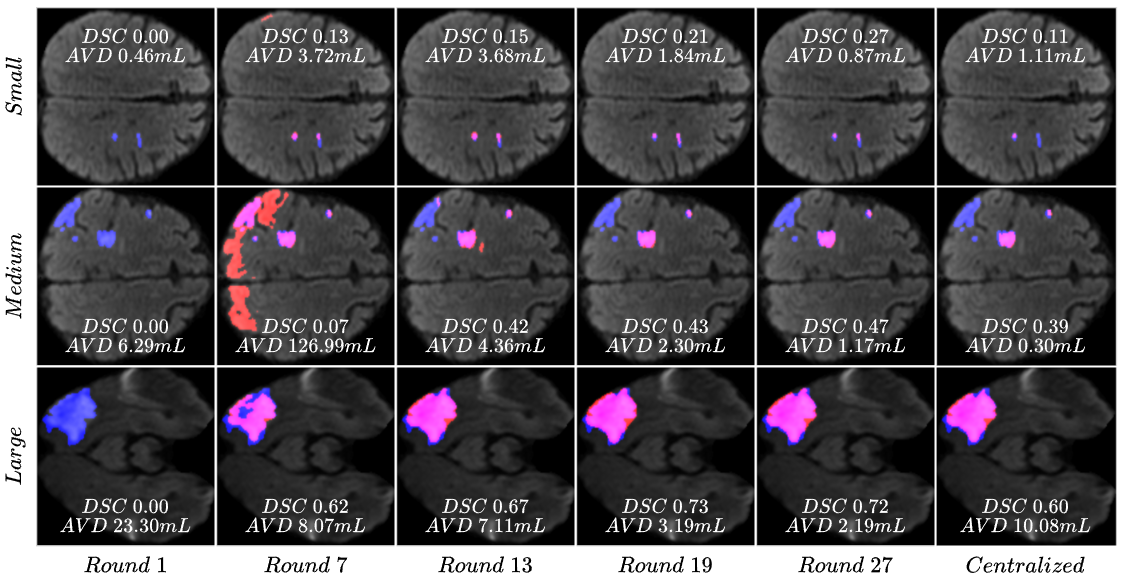}
    \caption{Progression of the FedAvg segmentation over DWI sequences for small, medium and large volume lesions over the rounds (blue color is the expert delineation, red is from models and pink is the intersection). The last column shows segmentation from the centralized baseline.}
    \label{fig:dwi_progression}
\end{figure*}

\section{Related works}
Diffusion-Weighted Magnetic Resonance Imaging (DWI-MRI) plays a crucial role in diagnosing ischemic stroke, evidencing intracellular cytotoxic and brain water accumulation as hyperintense signals in the images proportional to stroke advanced stages \cite{nukovic2023neuroimaging}. However, DWI sequences may report artifacts such as T2-shine-through or T2-blackout voxels, misleading the localization of ischemic lesions \cite{silvera2005spontaneous}. To complement such analysis, the Apparent Diffusion Coefficient (ADC) parametric map avoids these artifacts on DWI, showing the ischemic lesion as hypointense signals and allowing to confirm the localization and shape of flow obstruction, observed from DWI-B0 and DWI-B1000 sequences \cite{goyal2020challenging}. Findings observed in DWI are manually delineated and confirmed regarding the ADC map for quantifying the lesion volume, to support diagnosis and treatment planning, especially when symptom onset is unclear \cite{juan2022improving,nukovic2023neuroimaging}.

The lesion segmentation support can be obtained by using classic machine learning estimators, such as DecisionTree and RandomForest, or even by following computer vision strategies such as graph-cuts and standard statistical operators \cite{minaee2021image}. More recently, the predominant strategy has been to adjust deep UNet-like architectures through traditional centralized learning methods \cite{abbasi2023automatic,luo2024deep}. These approaches have explored feature integration of MRI multi-modal sequences with one or multiple processing branches, and with specialized modules to compute multi-scale and global relationships. For architectures with a single branch, Ashtari \textit{et al.} integrated DWI, ADC, and FLAIR images into a 3-channel volume using a single 3D UNet architecture that included a non-negative matrix factorization layer and a tensor-to-matrix shifted window operation to enhance multimodal lesion analysis \cite{ashtari2023factorizer}. Similarly, Goméz \textit{et. al.} integrated ADC and CT perfusion maps into a single 2D UNet network with cross-attention mechanisms that focus on the lesion shape through the convolutional saliency maps, adding skip connections to preserve the morphology of affected tissue. \cite{gomez2023deep}. For multi-branch architectures, Bal \textit{et al.} created a dual-path 3D convolutional network focused on local and non-local relationships within DWI and FLAIR images \cite{bal2024robust}. The local pathway captures fine-grained details within each slice, while the contextual pathway considers spatial dependencies between slices. Likewise, Goméz \textit{et. al.} proposed a dual-UNet multiparametric architecture for segmenting stroke lesions from ADC and DWI slices \cite{gomez2023ischemic}. The dual models extract relevant radiological findings individually and fuse them into a unified DWI-ADC embedding to enrich the reconstruction of embeddings into segmentation masks. Despite these models aim to exploit stroke imaging features, the effectiveness of these features is constrained by the size and diversity of the institution's dataset, potentially compromising the generalization capabilities of the models. This limitation may be critical in most of the clinical centers without labeled databases, which limits the applicability and support in constrained institutions due to the lack of variability \cite{luo2024deep,guan2024federated}.

\begin{figure*}
    \centering
    \includegraphics[width=0.95\textwidth]{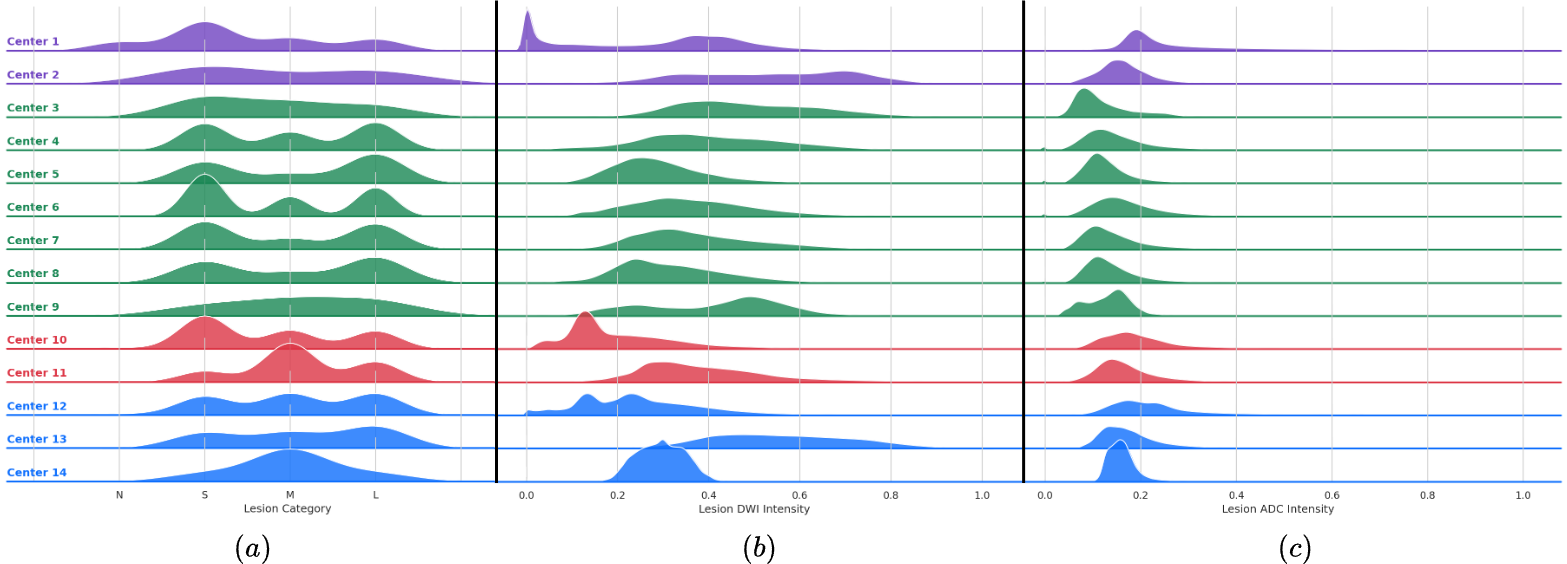}
    \caption{Distribution of the Federated Stroke Dataset constituted of 14 centers. (a) Volume categories for control (N), small (S), medium (M) and large (L) lesions among centers. (b) Lesion intensities from the DWI sequences. (c) Lesion intensities from the ADC parametric maps.}
    \label{fig:dist_heterogeneity}
\end{figure*}

\begin{table*}
    \caption{Healthcare centers properties and distribution, having the center category, voxel dimensions, scanner vendor and quantityof patients for train and test, respectively. \textit{NS} is for centers where the scanner vendor can not be defined, having one of the following: Philips (Achieva, Ingenia), Siemens Verio or Siemens MAGNETOM (Avanto, Aera)}
    \begin{tabularx}{\textwidth}{XXXXXX}
        \toprule
        \textbf{Center ID} & \textbf{Large Center} & \textbf{Voxels dimensions} $\left [ x, y, z\right ]$mm & \textbf{Scanner vendor} & \textbf{Scanner field strength} & \textbf{Train/Test Patients} \\ \midrule 
        \multicolumn{5}{l}{FOSCAL (180 patients)} \\ \midrule
        1 & $\checkmark$ & \small{[0.93, 0.93, 6.00]} & \small{Toshiba Vantage Titan} & 1.5T & 136 / 36 \\
        2 & \textit{x} & \small{[0.92, 0.92, 6.13]} & \small{General Electric Signa Explorer} & 1.5T & 0 / 8 \\ \midrule
        \multicolumn{5}{l}{SOOP (1451 patients) \cite{absher2024stroke}} \\ \midrule
        3 & \textit{x} & \small{[0.98, 0.98, 5.94]} & \small{General Electric Signa HDxt} & 1.5T & 0 / 16 \\
        4 & $\checkmark$ & \small{[0.89, 0.89, 6.00]} & \small{Philips Achieva} & 1.5T & 195 / 50 \\
        5 & $\checkmark$ & \small{[1.04, 1.04, 6.00]} & \small{Philips Achieva} & 3T & 96 / 25 \\
        6 & $\checkmark$ & \small{[1.04, 1.04, 6.00]} & \small{Philips Ingenia Ambition X} & 1.5T & 635 / 161 \\
        7 & $\checkmark$ & \small{[0.59, 0.59, 6.00]} & \small{Philips Ingenia Elition X} & 3T & 140 / 36 \\
        8 & $\checkmark$ & \small{[0.96, 0.96, 6.00]} & \small{Philips Intera} & 1.5T & 70 / 19 \\
        9 & \textit{x} & \small{[0.90, 0.90, 6.00]} & \small{Siemens Symphony} & 1.5T & 0 / 8 \\ \midrule
        \multicolumn{5}{l}{ISLES 22 (250 patients) \cite{petzsche2022isles}} \\ \midrule
        10 & $\checkmark$ & \small{[1.99, 1.99, 2.03]} & \small{\textit{NS}} & 1.5T or 3T & 157 / 41 \\
        11 & $\checkmark$ & \small{[1.51, 1.51, 4.80]} & \small{\textit{NS}} & 1.5T or 3T & 41 / 11 \\ \midrule
        \multicolumn{5}{l}{ISLES 24 (150 patients) \cite{riedel2024isles}} \\ \midrule
        12 & $\checkmark$ & \small{[0.43, 0.43, 2.00]} & \small{\textit{NS}} & 1.5T or 3T & 84 / 24 \\
        13 & $\checkmark$ & \small{[0.35, 0.35, 3.98]} & \small{\textit{NS}} & 1.5T or 3T & 31 / 10 \\
        14 & \textit{x} & \small{[0.49, 0.49, 0.80]} & \small{\textit{NS}} & 1.5T or 3T & 0 / 1 \\
        \bottomrule
        \label{tab:datasets_properties}
    \end{tabularx}
\end{table*}

The notable increase in the availability of public datasets for stroke has been crucial for the advancement and validation of computational algorithms \cite{liew2018large,liew2022large,maier_isles_2017,winzeck2018isles,petzsche2022isles,liu2023large,absher2024stroke}. For instance, Jeong \textit{et. al.} uses 3 different datasets with ADC and DWI sequences and implements two 3D UNet architectures for DWI and DWI-ADC input modalities. These models were evaluated by weighting the final segmentation mask using both outputs \cite{jeong2024robust}. Similarly, de la Rosa \textit{et. al.} implemented the 3 best segmentation architectures in ISLES 2022 challenge \cite{petzsche2022isles}, through majority voting between the outputs a final mask is generated \cite{de2024robust}. These algorithms leverage the heterogeneity of datasets evidencing new challenges to achieve generalization over limited and unseen hospitals. However, these ensemble approaches involve predictions from multiple models, demanding high computational, software and storage costs, creating an adoption barrier for smaller healthcare centers \cite{abbasi2023automatic, luo2024deep}.

Particularly for stroke segmentation, the literature on FL has been poorly explored. For instance, Otálora \textit{et. al.} implemented federated aggregation rules over ISLES 2018 dataset \cite{winzeck2018isles} using 94 CTP studies and simulating multiple centers by scanner vendor metadata \cite{otalora2022weighting}. However, this work contains annotations from only 1 expert and uses one simulated institution with only two scans, following a train-test evaluation. These limitations constrain the variability of radiologists and centers to forecast results in real-world scenarios. Alongside, Madrona \textit{et. al.} develops a FedAvg rule over a private dataset of 286 ADC images, simulating heterogeneity of centers by downsampling the images by varying the diffusion directions (\textit{b-vectors}) of neuroimages \cite{madrona2023federated}. This work emulates centers with different diffusion directions, being useful to complement representations from channels, but it is not a typical restriction among centers, and similarly to Otalora \textit{et. al.} the variability is hindered by the unique source of samples and simulated centers. On the other side, Rangel \textit{et. al.} implemented a FL scenario with two centers categories and datasets to simulate heterogeneity, enabling a collaborative platform for segmenting ischemic stroke lesions \cite{rangel2024federated}. However, this preliminary work only implemented one federated rule and required further comparison against a centralized baseline and other federated rules. Despite these approaches for ischemic stroke segmentation using FL, the inclusion of more datasets with enough variability of annotators, capture protocols and lesion sizes is critical to approximate simulated real-world scenarios, even surpassing detrimental properties from heterogeneous patient demographics, scanner vendors and lesion morphologies in the segmentation task.

\section{Materials and methods}
In this work, a standard FL framework was developed as an alternative methodology for segmenting ischemic stroke lesions, enabling a collaborative framework and evidencing generalization properties over several patients, scanner vendors and lesion morphologies and intensity representations. A federated stroke dataset with DWI and ADC neuroimages was built with 14 emulated healthcare centers preserving their heterogeneity attributes and divided between large (with model's training capabilities) and limited (out-of-distribution institutions). Besides, exhaustive validation shows reliable and uniform performance in limited centers without additional training regarding overlapping and clinical metrics in different lesion volume categories. The proposed approach is presented in figure \ref{fig:pipeline} where the large centers help in the construction of the federated model, while the federated model representation contributes in limited centers to support the segmentation task.

\subsection{Federated Stroke Dataset}
Despite the increasing number of public datasets for stroke segmentation, there is no federated-centered dataset yet. Commonly, the ischemic stroke datasets have been directed towards studying MRI neuroimages because of their greater sensitivity \textit{w.r.t} CT modalities \cite{powers2019guidelines,nukovic2023neuroimaging}. Hence, using the presence of DWI and ADC modalities a total of 3 public datasets were integrated in this work (ISLES 22 \cite{petzsche2022isles}, ISLES 24 \cite{riedel2024isles} and SOOP \cite{absher2024stroke}). Additionally, in cooperation with Fundacion Oftalmologica de Santander (FOSCAL) and FOSUNAB clinics in Colombia two private datasets were incorporated into the federation. The experimental protocol was approved by the ethics committees CEINCI-UIS at Universidad Industrial de Santander (approval number 4110, granted on February 10, 2023) and CEI-FOSCAL at Fundacion Oftalmologica de Santander (approval number 06075/2022, granted on April 29, 2022). Due to the retrospective nature of the study, the ethics committees CEINCI-UIS and CEI-FOSCAL waived the need for obtaining informed consent.

From such datasets, a total of 2031 studies with DWI sequences and ADC parametric maps were included and a framework has been established to recover and simulate centers with similar MRI scanner vendors and voxel dimensions between studies, obtaining a total of 14 emulated centers, detailed in table \ref{tab:datasets_properties}. In total 10 large centers were defined with enough patients to support federated model training, while 4 limited centers were set with a maximum of 16 patients for validating the generalization capabilities of AI models.

\subsection{Segmentation baseline model: nnUNet}
For each center, the segmentation task was performed using a nnUNet model \cite{isensee2021nnu} (formally defined as $\mathcal{F}^i \left (X_i; \theta \right ) = \hat{Y_i}$ where $X_i$ and $\hat{Y_i}$ represents the input MRI data and segmentation mask generated from the model, respectively, and $\theta_i$ the model parameters). The nnUNet model includes an UNet backbone together with pre- and post-processing automation techniques, consisting of a self-configuring strategy based on heuristic rules using dataset properties, such as intensity distribution, voxel sizes, image spacing, among others \cite{isensee2021nnu}. This self-configuration was carried out by sharing dataset properties from the 10 large centers to generate an optimal configuration for all models. A centralized baseline scenario was also included to compare performance against federated models by joining the 10 large centers' datasets. The four limited centers only received federated or central models to validate their generalization capabilities.

\subsection{Federated Learning on ischemic stroke}
In this work, a standard FL framework was developed fusing independent model representations into a single federated model (presented in figure \ref{fig:pipeline}). The models are shared at weight's instances instead of gradients to enhance patient privacy over the federated process \cite{rootes2021federated,yoo2022open}. The federated model architecture is initialized locally in each large center and adjusts the model concerning its own dataset. Then, the models are distributed to an aggregation server, which builds the federated model using equation \ref{eq: general_federated_agg}. Afterwards, the federated model is redistributed to centers and fine-tuned to complete a cycle of federation (\textit{a.k.a} federated round) and repeated until convergence \cite{joshi2022federated}. Formally, let $N$ independent clients' datasets $\left \{ \mathcal{D}_1, \mathcal{D}_2, \dots, \mathcal{D}_N \right \}$, with $\mathcal{D}_i = \left \{ X_i, Y_i \right \}$ being $X_i$ the input samples and $Y_i$ their expert annotated lesions. Also, the federated model is defined as $\mathcal{F}^{fed}\left (X_i; \theta \right ) = \hat{Y_i}$, with $\theta$ common training parameters that receive inputs $X_i$ and return predicted labels $\hat{Y_i}$ \cite{nguyen2022federated}. Then, the federated model is expressed as:

\begin{equation}
    \label{eq: general_federated_agg}
    \mathcal{F}^{fed}\left (X;  \theta \right ) = \sum_i^N \kappa \left ( i \right ) \mathcal{F}^i \left (X;  \theta_i \right )
\end{equation}

Commonly, $\mathcal{F}^{fed}\left ( X_i; \theta \right ) = \hat{Y_i}$ is the sum of local models $\mathcal{F}^i \left ( X_i;  \theta_i \right ) = \hat{Y_i}$, with parameters $\theta_i$ \cite{sharma2023comprehensive}. The $\kappa \left ( i \right )$ is a mapping function (\textit{a.k.a} aggregation rule) to assign a value dependent on the client $i$ (healthcare center), and the following aggregation rules were implemented:

\begin{itemize}
    \item \textbf{FedAvg rule}. $\kappa \left ( i \right )$ is defined as $\kappa \left ( i \right ) = \lvert \mathcal{D}_i \rvert / \sum^{N}_{j} \lvert \mathcal{D}_j \rvert$, having $\lvert \mathcal{D}_i \rvert$ as the cardinality of the dataset or number of samples in client $i$. This rule works under the assumption that larger datasets contribute more variability and generalization capabilities inside the centers, assigning more weight ($\kappa \left ( i \right )$ value) to bigger centers \cite{mcmahan2017communication}.
    \item \textbf{Equally weighted (VanillaAvg) rule}. In such case $\kappa \left ( i \right )$ is equal for all centers ($\kappa \left ( i \right ) = 1 / N$), assuming that the variability of each center is equally relevant \cite{sharma2023comprehensive}.
    \item \textbf{Beta Weighting rule} \cite{otalora2022weighting}. In this case $\kappa \left ( i \right ) = \mathcal{W} \left ( i \right ) / \sum^{N}_{j} \mathcal{W} \left ( j \right )$ where $\mathcal{W} \left ( i \right ) = \left ( 1 - \beta \right ) / \left ( 1 - \beta^{\lvert \mathcal{D}_i \rvert} \right )$ with $\beta$ between $\left [ 0, 1\right ]$. This rule works under the assumption that smaller centers contributes equally against bigger centers, but are limited in the frequency of samples, assigning more weight ($\kappa \left ( i \right )$ value) to smaller centers to balance the contributions between smaller and bigger centers \cite{cui2019class}.
    \item \textbf{Softmax aggregation rule} \cite{otalora2022weighting}. Here the $\kappa \left ( i \right )$ is defined as: $e^{\lvert \mathcal{D}_i \rvert} / \sum^{N}_{j} e^{\lvert \mathcal{D}_j \rvert}$. Although this rule works similarly to FedAvg, the $\kappa \left ( i \right )$ values are assigned in an exponential function, weighting even more bigger centers and ignoring the smaller ones. 
    \item \textbf{FedProx rule.} $\kappa \left ( i \right )$ is equal as FedAvg but the training loss of local models adds a regularization term as $\mathscr{L}_{total}^i = \mathscr{L}^i \left ( Y_i, \hat{Y_i}\right ) + \frac{\mu}{2} || \mathcal{F}^{i}\left ( X_i; \theta \right ) - \mathcal{F}^{fed}\left ( X_i; \theta \right ) ||^2$ where the federated model weights constraint the heterogeneity inside the local models to regularize the updates $\mu$ quantity. This rule works under the assumption that the federated model is totally generalizable, then the local models should be closer to the federated representation \cite{li2020federated}.
\end{itemize}

An advantage compared to ensemble approaches is that these aggregation rules build a feasible computational solution for small clinical institutions, by generating a single model that contains the variability learned from participant centers to generate ischemic stroke segmentation masks. This federated model can be employed in new centers to support stroke lesion characterization with limited training samples or expert annotations (\textit{a.k.a} out of distribution centers). At the same time, the participating centers benefit from the variability learned from other centers without compromising data privacy and security generating more robust and reliable segmentations.

\begin{table*}
    \caption{Segmentation results regarding 10 large and 4 limited centers from the federated stroke datasets. The mean relative error for patients and metrics is included indicating the average deviation rate from a perfect segmentation regarding the overlapping and clinical metrics. The first and second best results are bolded and underlined, respectively.}
    \begin{center}
        \begin{tabular}{lccccc}
            \toprule
             & \textbf{PRE} $\downarrow$ & \textbf{DSC} $\uparrow$ & \textbf{AVD [mL]} $\downarrow$ & \textbf{ALD} $\downarrow$ & \textbf{LF1} $\uparrow$ \\ \midrule
            \multicolumn{6}{l}{Large centers} \\ \midrule
            Centralized & $\underline{0.58 \pm 0.18}$ & $\underline{0.71 \pm 0.25}$ & $5.57 \pm 24.74$ & $\underline{2.33 \pm 3.87}$ & $\underline{0.69 \pm 0.26}$ \\
            FedAvg & $\mathbf{0.57 \pm 0.19}$ & $\mathbf{0.72 \pm 0.24}$ & $\mathbf{5.35 \pm 23.50}$ & $\mathbf{2.15 \pm 3.63}$ & $\mathbf{0.71 \pm 0.26}$ \\
            VanillaAvg & $\underline{0.58 \pm 0.18}$ & $0.68 \pm 0.25$ & $6.85 \pm 27.68$ & $\underline{2.33 \pm 3.48}$ & $\underline{0.69 \pm 0.26}$ \\
            Beta Weighting & $0.59 \pm 0.20$ & $0.66 \pm 0.27$ & $7.68 \pm 29.71$ & $2.34 \pm 4.04$ & $\underline{0.69 \pm 0.28}$ \\
            Softmax & $0.59 \pm 0.19$ & $0.69 \pm 0.26$ & $6.40 \pm 25.46$ & $2.56 \pm 4.38$ & $\underline{0.69 \pm 0.28}$ \\
            FedProx & $0.59 \pm 0.19$ & $0.69 \pm 0.25$ & $\underline{5.48 \pm 24.34}$ & $2.37 \pm 3.97$ & $0.68 \pm 0.26$ \\ \midrule
            \multicolumn{6}{l}{Limited centers} \\ \midrule
            Centralized & $\mathbf{0.58 \pm 0.23}$ & $\underline{0.63 \pm 0.30}$ & $\mathbf{4.21 \pm 8.12}$ & $\mathbf{2.09 \pm 2.93}$ & $\mathbf{0.66 \pm 0.35}$ \\
            FedAvg & $\mathbf{0.58 \pm 0.20}$ & $\mathbf{0.64 \pm 0.29}$ & $\underline{4.44 \pm 8.74}$ & $2.24 \pm 3.29$ & $\underline{0.63 \pm 0.33}$ \\
            VanillaAvg & $0.61 \pm 0.22$ & $0.61 \pm 0.29$ & $4.56 \pm 8.01$ & $2.91 \pm 4.35$ & $0.59 \pm 0.36$ \\
            Beta Weighting & $0.62 \pm 0.23$ & $0.58 \pm 0.30$ & $5.54 \pm 9.49$ & $2.52 \pm 3.48$ & $0.61 \pm 0.34$ \\
            Softmax & $0.61 \pm 0.25$ & $0.59 \pm 0.31$ & $4.94 \pm 7.95$ & $2.27 \pm 2.78$ & $0.61 \pm 0.32$ \\
            FedProx & $\underline{0.60 \pm 0.21}$ & $\underline{0.63 \pm 0.29}$ & $4.96 \pm 9.20$ & $\underline{2.21 \pm 3.21}$ & $0.62 \pm 0.33$ \\
            \bottomrule
            \label{tab:results_per_center_category}
        \end{tabular}
    \end{center}
\end{table*}

\section{Experimental Setup}
\subsection{Federated validation}
An exhaustive validation was developed considering different institutions, lesion categories, overlapping and clinical metrics. The overlapping metrics were focused on delineation similarity (Dice Score or DSC) and localization of the lesion (Lesion F1 Score or LF1). As for clinical metrics, the absolute volume difference (AVD) and absolute lesion difference (ALD) were considered to estimate over or under delineation of the lesion and a similar number of detected lesions, respectively. The lesions were categorized in Control (N, volume of $0 \; mL$), Small (S, volume $\leq 5 \; mL$), Medium (M, volume $> 5 \; mL \; \& \; \leq 20 \; mL$) and Large (L, volume $> 20 \; mL$), where small lesions requires a higher detection metrics (\textit{e.g} LF1 and ALD) due to greater salvageable brain tissue during stroke, while large lesions requires higher delineation metrics (\textit{e.g} DSC and AVD) because the functional outcome and treatment prediction benefits from accurate estimations \cite{sperber2023stroke, de2024robust}. The metrics were estimated for each patient, where large centers carry out a stratified random split of 80/20, and limited centers are test-only cases to validate generalization capabilities. To determine the best model, a ranking system was implemented over each metric at patient level and averaged to obtain the model's final position \cite{maier_isles_2017, winzeck2018isles}. However, considering only participant models introduce favorability bias and ignore metrics limitations for different lesion categories requiring a proximity term to the expert delineation (\textit{i.e.} from 20 models, AVD metric for the segmentation in a small lesion are between 10 and 50 mL, returning first position to the model with 10 mL in AVD but ignoring the overestimation error in the lesion) \cite{reinke2024understanding}.

Therefore, in this work the ranking system is based on the relative error from radiologists' delineation ($\delta X = \lvert \; X_{model} - X_{expert} \rvert / X_{expert}$). The position is the mean Patients Relative Error (PRE) defined as $PRE = \sum_{pat \in P} \sum_{M_i \in pat} \delta M_i / \left ( \lvert M_i \rvert \cdot \lvert P \rvert \right )$ where $\lvert P \rvert$ is the number of patients, $M_i$ are the metrics considered for the ranking (DSC, AVD, ALD and LF1), and $\lvert M_i \rvert$ is the quantity of metrics. This metric avoids classical ranking problems, helps to untie models with similar performances, and reports how far the results are from being exactly as the expert segmentation. It is important to note that relative errors are clipped between 0 and 1, where 1 indicates that the model is generating misguided segmentations. Additionally, a centralized baseline is included in the ranking as the upper limit of the metrics to compare against federated rules.

\subsection{Backbone configuration}
The nnUNet architecture of each institution was adapted to receive 2D DWI and ADC concatenated slices in channel dimension. The main computation block operations for each processing level were defined as 2 consecutive sets of operations consisting of a $3 \times 3$ convolution, leaky ReLU and instance normalization, followed by pooling or transposed convolution which halves or doubles $2 \times 2$ the features depending on if it's the encoder or decoder module, respectively. In total, the architecture has six processing levels, where the number of filters starts at 32 and doubles at each subsequent level, being the max up to 512 features. Regarding the federative training process, the local models have a fixed amount of 20 epochs and a total of 30 federated rounds to reach 600 training epochs per model among healthcare centers. 

The nnUNet self-configuration was carried out on the grouped datasets from 10 large centers to generate the optimal configurations for the central baseline and federated segmentation models. Furthermore, this self-configuration established a standard preprocessing pipeline for large and limited centers to enable a fair comparison between models. Lastly some agnostic preprocessing functions were included in the pipeline, such as skull stripping with synthstrip \cite{hoopes2022synthstrip}, filtering outside brain segmentations (mask filtering), registering modalities with ANTs \cite{avants2009advanced}, brain-focused crop and $1 \; mm^2$ isotropic voxel resampling over the slices to help the model learn relevant features of ischemic stroke over the brain only.

\section{Evaluations and results}

\begin{figure}
    \centering
    \includegraphics[width=0.5\textwidth]{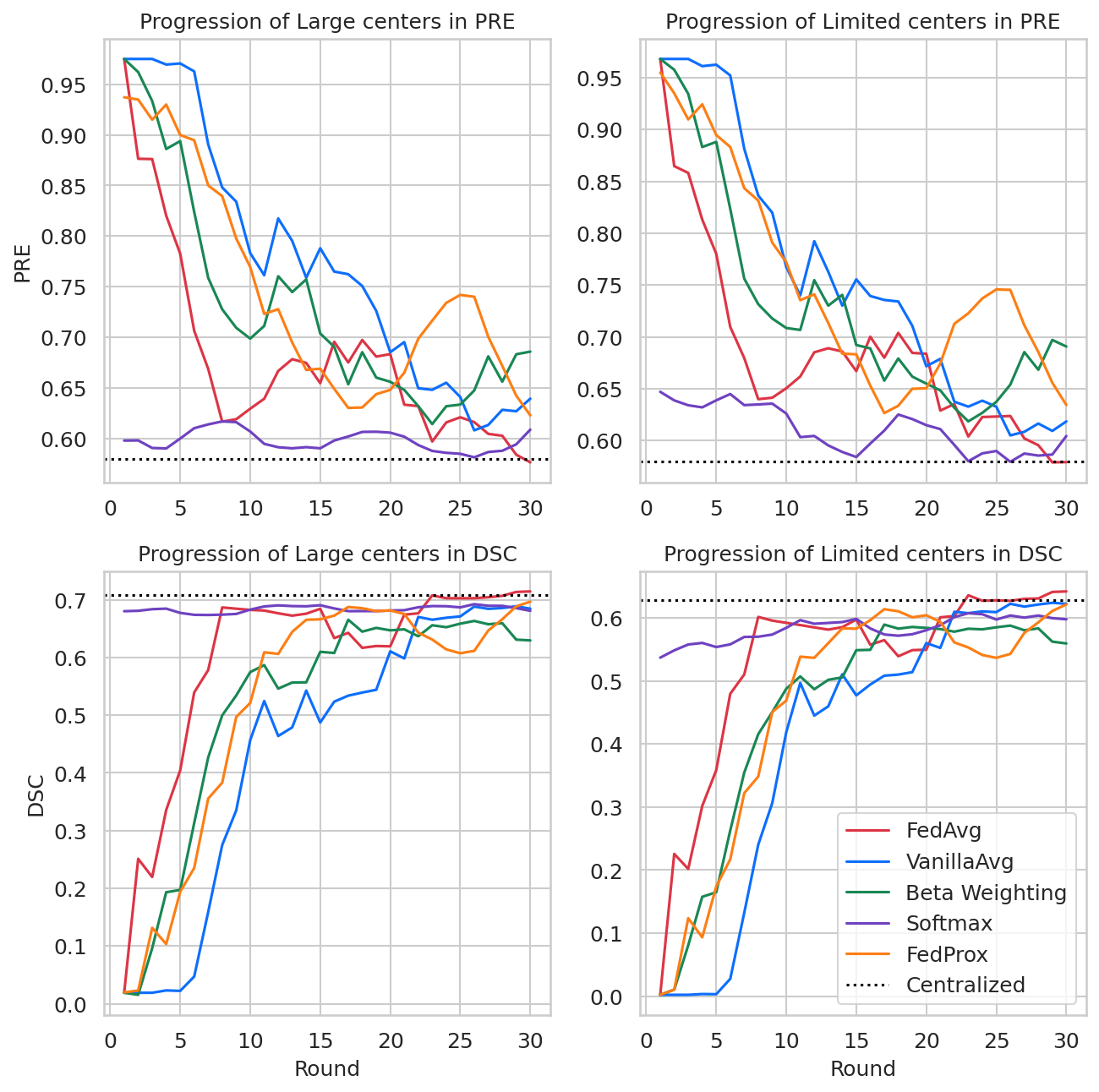}
    \caption{Progression of the 5 federated aggregation rules over the rounds. The first row is for Patients Relative Error (PRE) of models and the second is the Dice Score (DSC). The columns are for large and limited healthcare centers, respectively.}
    \label{fig:progression}
\end{figure}

A main interest in this work was to develop federated learning for ischemic stroke segmentation, as an alternative to support independent centers with remarked heterogeneity, regarding lesion volume categories and DWI intensities. Figure \ref{fig:dist_heterogeneity} summarizes per-centre data distributions according to category, DWI and ADC intensities for lesion observations. As expected, the remarked heterogeneity between distributions is observed among healthcare centers. While centers 11 and 14 reported prevalent medium lesions, centers 1, 6 and 10 had a major occurrence of small lesions. The intensities of ischemic stroke lesions were normalized \textit{w.r.t} brain intensities to set a common domain for all centers. Even the proportion and DWI lesion intensity distributions showed evident heterogeneity between centers. Moreover, table \ref{tab:datasets_properties} reports voxel dimension heterogeneity among centers. Contrary to these findings, ADC distributions (figure \ref{fig:dist_heterogeneity}c) are consistent among centers to confirm the flow obstruction with hypointense signals (around $620 \times 10^{-6} \; mm^2/s$ \cite{pistocchi2022mri}).

A federated training was carried out using 10 large centers, following 5 aggregation rules: FedAvg, VanillaAvg, Beta Weighting, Softmax and FedProx. These rules are compared against a centralized baseline trained on the joint dataset from large centers. Table \ref{tab:results_per_center_category} reports the segmentation results for large and limited centers. Surprisingly, FedAvg surpassed the performance, while VanillaAvg was similar to the centralized baseline, as indicated by the PRE ranking metric for large centers. Regarding the overlapping metrics, FedAvg obtained 1\% and 2\% more DSC and LF1 than the centralized baseline, indicating more reliable segmentation and detection of ischemic lesions in DWI and ADC. In terms of clinical metrics (AVD and ALD), the FedAvg doesn't over or under estimate the volume and number of lesions by 0.22 mL and 0.18 lesions, respectively, compared to the centralized approach. Interestingly, the FedProx aggregation rule obtained the second position in AVD with underestimated volumes compared to the centralized baseline, showing the potential use of different rules for different tasks in ischemic stroke diagnosis.

Regarding the limited centers, table \ref{tab:results_per_center_category} shows a tie between the centralized and FedAvg models with a PRE of 0.58, where the centralized model slightly outperforms the federated solutions regarding AVD, ALD and LF1 metrics with 0.21 mL, 0.12 lesions and 3\% of difference, respectively. The FedAvg model obtained better performance in DSC with 1\% more. Additionally, the analogous performance between FedAvg and centralized models in PRE metric evidences the stronghold of relative error for ranking the models by avoiding the favorability bias between metrics. In other words, the FedAvg segmentations for each patient present more confidence regarding delineation properties, while lacking in computing accurate volume or number of lesions in DWI and ADC studies due to over/under estimating the ischemic stroke segmentations. Therefore, FedAvg evidences potential use in limited centers without enough training capabilities.

In figure \ref{fig:progression} is shown the progression of the aggregation rules. At the top PRE progression for large and limited centers (left and right graphs, respectively), where FedAvg reported faster convergence, even surpassing the centralized approach. Such a fact evidences the federative capabilities for heterogeneous scenarios, while generating robust segmentations in out-of-distribution centers. Similarly, at the bottom the DSC of federated approaches reaches the upper bound (centralized model) for large and limited centers, where the FedAvg also presents faster convergence.

To complement the validation, table \ref{tab:results_per_lesion_category} reports the analysis per-volume category. The performance for centers that participated in the federated training setting (large centers) evidenced that the FedAvg approach outperformed the centralized baseline in small ($\leq$ 5.0 mL), medium ($>$ 5.0 mL \& $\leq$ 20.0 mL) and large ($>$ 20.0 mL) lesion volumes around 1.33\% less PRE, with larger lesions having the best delineation (DSC of 83\%) and small lesions having the best detection (LF1 of 72\%). Interestingly, the performance between the lesion categories is consistent on PRE, indicating that the proportion of patients manifests a common management for different lesions regarding the segmentation task. On the other hand, for limited centers (out-of-distribution institutions) the FedAvg approach surpassed the performance of the centralized baseline regarding small and medium lesion volumes around 4\% less PRE. Curiously, the DSC and AVD for large lesion volumes are better for the FedAvg model by 2\% and 0.02 mL than the centralized baseline, pointing out that the influence of large lesions misleads in the final segmentation results because of their high difference in volume (\textit{i.e.} the centralized PRE is the lowest for large but surpass 60\% of error for small and medium volumes). Additionally, the performance of FedAvg between participating and out-of-distribution centers is similar, indicating the generalization capabilities for new healthcare centers that require a segmentation model. In figure \ref{fig:dwi_progression} the segmentation for small, medium and large lesions is shown, complementing the generalization properties of FedAvg against the centralized approach in out-of-distribution centers, regarding the segmentation task.

\begin{table*}
    \caption{Results obtained for each lesion volume category between centralized and FedAvg over the 10 institutions that participated in the federated training (Large centers) and 4 beneficiary institutions from the federated model (Limited centers). The highest values are bolded for each metric.}
    \begin{center}
        \begin{tabular}{lcccc}
            \toprule
             & \multicolumn{2}{c}{\textbf{Large centers}} & \multicolumn{2}{c}{\textbf{Limited centers}} \\ \cmidrule(lr){2-3} \cmidrule(lr){4-5}
             & Centralized & FedAvg & Centralized & FedAvg \\ \midrule
            \multicolumn{5}{l}{Small lesions ($\leq$ 5.0 mL)} \\ \midrule
            \textbf{PRE} $\downarrow$ & $0.59 \pm 0.21$ & $\mathbf{0.57 \pm 0.21}$ & $0.60 \pm 0.29$ & $\mathbf{0.58 \pm 0.27}$ \\
            \textbf{DSC} $\uparrow$ & $0.61 \pm 0.26$ & $\mathbf{0.63 \pm 0.24}$ & $0.44 \pm 0.37$ & $\mathbf{0.46 \pm 0.37}$ \\
            \textbf{AVD [mL]} $\downarrow$ & $\mathbf{0.58 \pm 0.67}$ & $0.64 \pm 1.00$ & $1.39 \pm 3.06$ & $\mathbf{1.21 \pm 2.84}$ \\
            \textbf{ALD} $\downarrow$ & $1.66 \pm 2.49$ & $\mathbf{1.42 \pm 2.05}$ & $\mathbf{1.23 \pm 1.64}$ & $1.46 \pm 2.07$ \\
            \textbf{LF1} $\uparrow$ & $0.69 \pm 0.29$ & $\mathbf{0.72 \pm 0.29}$ & $\mathbf{0.62 \pm 0.43}$ & $\mathbf{0.62 \pm 0.42}$ \\ \midrule
            \multicolumn{5}{l}{Medium lesions ($>$ 5.0 mL \& $\leq$ 20.0 mL)} \\ \midrule
            \textbf{PRE} $\downarrow$ & $0.59 \pm 0.16$ & $\mathbf{0.58 \pm 0.16}$ & $0.61 \pm 0.16$ & $\mathbf{0.55 \pm 0.16}$ \\
            \textbf{DSC} $\uparrow$ & $\mathbf{0.73 \pm 0.21}$ & $\mathbf{0.73 \pm 0.21}$ & $\mathbf{0.71 \pm 0.15}$ & $\mathbf{0.71 \pm 0.13}$ \\
            \textbf{AVD [mL]} $\downarrow$ & $2.93 \pm 2.93$ & $\mathbf{2.78 \pm 2.76}$ & $\mathbf{3.05 \pm 2.78}$ & $4.19 \pm 3.88$ \\
            \textbf{ALD} $\downarrow$ & $2.64 \pm 4.00$ & $\mathbf{2.58 \pm 3.99}$ & $2.30 \pm 1.95$ & $\mathbf{2.00 \pm 2.26}$ \\
            \textbf{LF1} $\uparrow$ & $0.70 \pm 0.24$ & $\mathbf{0.71 \pm 0.24}$ & $0.67 \pm 0.28$ & $\mathbf{0.68 \pm 0.28}$ \\ \midrule
            \multicolumn{5}{l}{Large lesions ($>$ 20.0 mL)} \\ \midrule
            \textbf{PRE} $\downarrow$ & $0.57 \pm 0.13$ & $\mathbf{0.56 \pm 0.14}$ & $\mathbf{0.53 \pm 0.23}$ & $0.62 \pm 0.15$ \\
            \textbf{DSC} $\uparrow$ & $\mathbf{0.83 \pm 0.17}$ & $\mathbf{0.83 \pm 0.16}$ & $0.80 \pm 0.16$ & $\mathbf{0.82 \pm 0.14}$ \\
            \textbf{AVD [mL]} $\downarrow$ & $13.16 \pm 40.88$ & $\mathbf{12.61 \pm 38.83}$ & $9.04 \pm 13.23$ & $\mathbf{8.90 \pm 14.44}$ \\
            \textbf{ALD} $\downarrow$ & $2.99 \pm 5.00$ & $\mathbf{2.82 \pm 4.70}$ & $\mathbf{3.00 \pm 4.59}$ & $3.50 \pm 4.99$ \\
            \textbf{LF1} $\uparrow$ & $0.69 \pm 0.22$ & $\mathbf{0.71 \pm 0.21}$ & $\mathbf{0.69 \pm 0.32}$ & $0.61 \pm 0.28$ \\
            \bottomrule
            \label{tab:results_per_lesion_category}
        \end{tabular}
    \end{center}
\end{table*}

\section{Discussion}
This work explored the key advantages and limitations of FL to address the limitations of data scarcity at each institution in the context of ischemic stroke lesion segmentation over DWI sequences and ADC parametric maps. A total of 5 state-of-the-art merging rules were herein considered in a standard federative scheme with a central joint model. The federated models were assembled by fusing independent deep learning models over heterogeneous healthcare centers, considering their shift-institution characteristics such as lesions proportions, intensities, scanner vendors and expert delineations. The aggregation rules were exhaustively validated regarding overlapping and clinical metrics, and ranked to determine the best approach for centers with and without training capabilities, conserving their heterogeneity properties to approximate real-world scenarios. This study included 2031 patients from 14 healthcare centers divided into large (where the model learns to segment) and out-of-distribution institutions (limited centers), and the lesion volumes were categorized as control (N), small (S), medium (M) and large (L) to complement the validation of the models. An exhaustive experimental setup was conducted to measure the overlap between the model's segmentations and the expert delineations, the federative generalization capabilities regarding limited centers in heterogeneous capture settings, scanner vendors and population demographics. The FedAvg aggregation rule achieved a general DSC of $0.71 \pm 0.24$, AVD of $5.29 \pm 22.74$, ALD of $2.16 \pm 3.60$ and LF1 of $0.70 \pm 0.26$ over the complete dataset pool, even surpassing the centralized baseline. Besides, the federated model evidenced robust performance among lesion categories, with detection capabilities in small lesions and improved delineations over large lesions.

From reported results the federative schemes could cover the dependency of data diversity and heterogeneous distributions, showing a competitive performance of DSC and LF1 around 70\% for healthcare centers that participate in the training of models and out-of-distribution samples, similarly to other approaches \cite{abbasi2023automatic,luo2024deep,gomez2023deep,gomez2023ischemic,ashtari2023factorizer,bal2024robust,jeong2024robust,de2024robust,otalora2022weighting,madrona2023federated}. Additionally, evaluating the PRE metric from the centralized baseline unveils a different perspective reporting a 58\% of patients (239 of 413 patients) with limited performance regarding the overlapping or clinical metrics, while the FedAvg model performs better with 4 patients less than the centralized baseline. Curiously, the PRE metric remains with similar performance in limited centers, suggesting consistent behaviour for different patient cohorts.

In figure \ref{fig:progression} the FedAvg approach evidenced the fastest convergence against all the other rules, even surpassing the upper threshold from the centralized baseline. Interestingly, the DSC metric progression showed a linear growth rate until round 25, then federated models struggled to reach centralized baseline performance. This slower rate may be associated with the implicit nature of federated merging rules, based on $\kappa \left ( i \right )$ weight functions, requiring patients' quantity instead of dataset characteristics, resulting in similar performance as centralized models \cite{nguyen2022federated, yoo2022open, sharma2023comprehensive}. To complement the obtained results, an exhaustive validation over the stratified lesions volume categories was calculated in table \ref{tab:results_per_lesion_category}. The FedAvg surpassed the centralized baseline in all overlapping and clinical metrics for large centers, evidenced by 1.4\% less PRE value (approximately 6 fewer patients with erroneous segmentations). Regarding small lesions (S, $\leq 5.0\text{mL}$), the FedAvg resulted in 2\% less PRE value than centralized model (around 4 fewer erroneous masks) in large and limited centers. Moreover, the federated model presented a detrimental effect on DSC performance for unseen patients, associated with an approximate 11.4\% of lesions being over/under delineated from radiologists' annotation; however, for this kind of lesions takes more relevance the detection instead of delineation tasks where the minor performance loss is evidenced in the LF1 metric \cite{sperber2023stroke, de2024robust}. Similarly, the FedAvg rule outperforms the centralized approach in medium lesions (M, $> 5.0\text{mL} \; \& \; \leq 20.0\text{mL}$), with 6\% less PRE value and consistent performance between seen and unseen populations. Contrary to these findings, large lesions (L, $> 20.0\text{mL}$) presented better performance only in DSC and AVD metrics, resulting in a higher PRE value than centralized baseline. However, this kind of lesions requires improved delineations in terms of DSC and AVD to accurately estimate the volume and deliver proper treatment decisions and outcome prognosis of patients \cite{neumann2009interrater}. Despite the competitive performance of the centralized baseline, the federated model evidenced a constant performance for small, medium and large lesions compared to experts' annotations. Remarkably, the FedAvg approach evidenced a consistent performance per-centre and per-lesion categories (tables \ref{tab:results_per_center_category} and \ref{tab:results_per_lesion_category}), contributing to the implicit generalizable properties of FL against centralized methodologies (shown in figure \ref{fig:dwi_progression}).

\section{Conclusions}
The study confirmed that Federated Learning is a promising alternative for supporting ischemic stroke segmentation in diverse real-world scenarios, tackling variability in patient demographics, expert annotations, scanner vendors, and lesion morphology. The framework also delivers functional segmentation models to centers without training or fine-tuning capabilities, all while complying with patient privacy regulations. The results showed generalizable capabilities with constant performance across patient demographics and lesion volume categories, obtaining a general DSC of $0.71 \pm 0.24$, AVD of $5.29 \pm 22.74$, ALD of $2.16 \pm 3.60$ and LF1 of $0.70 \pm 0.26$ over all centers, outperforming both the centralized and other federated rules. Besides, federated models evidenced reliable and uniform performance in out-of-distribution centers (with DSC of $0.64 \pm 0.29$ and AVD of $4.44 \pm 8.74$ without any additional training). However, ischemic stroke lesion segmentation remains limited due to the high variability between studies and patients. Future works include the study of new aggregation rules based on center and dataset latent properties, while also exploring alternative methods to improve local representations without requiring a complete transformation of deep segmentation models inside the centers.

\section*{Code Availability}
The code that supports the results, graphs and metrics obtained are available online in \url{https://gitlab.com/bivl2ab/research/publications/2025_fed-ischemic-stroke}






\bibliographystyle{elsarticle-num}
\bibliography{references}

\end{document}